\documentclass[preprint,superscriptaddress,nofootinbib]{revtex4}
\usepackage[dvips]{graphicx}
\usepackage[usenames]{color}
\usepackage{setspace}
\usepackage{amsmath}
\usepackage{amssymb}
\usepackage{amsthm}
\usepackage{bm}
\usepackage{hyperref}

\begin{document}

 \preprint{UT--HET--094}
 \preprint{EPHOU--15--008}
 \preprint{FTUV--15--8555}
 \preprint{IFIC--15--15}


 \title{Phenomenological constraints on light mixed sneutrino dark matter scenarios}
 \author{Mitsuru Kakizaki}
 \email{kakizaki@sci.u-toyama.ac.jp}
 \affiliation{
 Department of Physics,
 University of Toyama, Toyama 930-8555, Japan
 }
 \author{Eun-Kyung Park}
 \email{epark@particle.sci.hokudai.ac.jp}
 \affiliation{
 Department of Physics,
 Hokkaido University, Sapporo 060-0810, Japan
 }
 \author{Jae-hyeon Park}
 \email{jae.park@uv.es}
 \affiliation{
 Departament de F\'{i}sica Te\`{o}rica and IFIC,
 Universitat de Val\`{e}ncia-CSIC,
 46100, Burjassot, Spain
 }
 \author{Akiteru Santa} 
 \email{santa@jodo.sci.u-toyama.ac.jp}
 \affiliation{
 Department of Physics,
 University of Toyama, Toyama 930-8555, Japan
 }
 \begin{abstract}
   In supersymmetric models with Dirac neutrinos, the lightest
   sneutrino can be a good thermal dark matter candidate when
   the soft sneutrino trilinear parameter is large. In this paper, 
   we focus on scenarios where the mass of the mixed sneutrino LSP 
   is of the order of GeV so the sneutrino dark matter is still viable
   complying with
   the limits by current and near future direct detection experiments.    
   We investigate phenomenological constraints in the parameter
   space of the models, as well as the vacuum stability bound.  
   Finally, we show that the allowed regions can be explored by measuring
   Higgs boson properties at future collider experiments.
 \end{abstract}
\maketitle

\cleardoublepage
\pagenumbering{arabic}

\section{Introduction}

On July 4th, 2012, the ATLAS and CMS collaborations of the CERN Large Hadron 
Collider (LHC) announced the discovery of a new particle with a mass 
of 125 GeV \cite{LHC}.
The spin and parity properties of the new particle as well as its
couplings to Standard Model (SM) particles have been investigated,
and proven to be consistent with the prediction of the SM\@.
The SM has been established as a low energy effective theory
that explains phenomena at energy scales below ${\cal O}(100)$ GeV\@.

Although the SM is extraordinarily successful, there are still
unresolved problems.  The observation of neutrino oscillations reveals
that neutrinos must have finite masses and contradicts the SM, where
the neutrinos are massless \cite{SK}.  Cosmological observations
precisely determine the energy density of dark matter (DM) in the universe
while there is no candidate particle that can fulfill the dark matter
abundance in the SM \cite{WMAP,Planck:2013pxb}.
From the theoretical viewpoint, in order to explain the observed Higgs
boson mass in the framework of the SM an unnaturally huge fine-tuning
between its bare mass squared and contributions from radiative
corrections is required.
We are obliged to construct a more fundamental theory beyond the SM
to tackle these difficulties.

The problems mentioned above are solvable in supersymmetric (SUSY)
extensions with right-handed neutrino chiral supermultiplets 
\cite{HallMoroi:1997,ArkaniHamed:2000bq,BorzumatiNomura:2000,Alan:2004,AsakaMoroi:2005,
Arina:2007,Thomas:2008,Deppisch,CerdenoSeto:2009,Belanger:2010cd,Belanger:2011,
KhalilOkada:2011,ChoiSeto:2012,Dumont:2012ee,Arina:2014,Arina:2015uea}. 
The couplings of the right-handed neutrinos to the left-handed
counterparts provide a source of the observed neutrino masses, which
are either Dirac- or Majorana-type.  The hierarchy problem is avoided
by introducing SUSY: The quadratically divergent SM contributions to
the Higgs boson mass squared are canceled out by those from diagrams
involving superparticles whose spins differ from their SM counterparts
by half a unit.  It is intriguing that a viable candidate for dark matter
other than conventional ones is automatically introduced as a
by-product in this framework: The lightest sneutrino which is mainly
m of the right-handed component.
When such a sneutrino is the lightest SUSY particle
(LSP), the observed dark matter abundance can be explained
while satisfying other experimental constraints,
in sharp contrast to left-handed sneutrino LSP scenarios which are
excluded by the data of direct detection of dark matter.
In particular, 
SUSY scenarios with Dirac neutrinos and large SUSY breaking 
sneutrino trilinear parameters can provide a viable left-right mixed sneutrino 
dark matter candidate \cite{Arina:2007,
Thomas:2008,Belanger:2011,Arina:2014,Alan:2004,Belanger:2010cd,Dumont:2012ee,Arina:2015uea}.
Sneutrino trilinear parameters of the order of other soft SUSY breaking
masses can be naturally realized in models
where $F$-term SUSY breaking is responsible for the smallness
of the neutrino Yukawa couplings and induce large mixings
between the left- and right-handed sneutrino states \cite{ArkaniHamed:2000bq} .
Due to the large sneutrino trilinear coupling, the lightest mixed sneutrino 
behaves as a weakly interacting massive particle (WIMP) and 
its thermal relic abundance falls in the cosmological dark matter abundance.
So far, such mixed sneutrino WIMP scenarios have been 
screened in the light of experimental results.
If the mixed sneutrino mass is of the order of 100 GeV, 
its thermal relic abundance can account for the observed dark matter abundance
without contradicting experimental constraints.
On the other hand, when the mass of the mixed
sneutrino is smaller than half the mass of the discovered Higgs boson,
its invisible decay rate is significantly enhanced.
It has been shown that such a light sneutrino dark matter scenario
is excluded in the light of the LHC results
if the gaugino mass universality is imposed \cite{Dumont:2012ee}.

In this paper, we explore the GeV-mass mixed sneutrino scenarios
without gaugino mass universality.
We show that when the lightest neutralino mass is of the order 
of the mixed sneutrino mass, 
the thermal relic abundance of the mixed sneutrino coincides with
the observed dark matte abundance.
It should be emphasized that the large sneutrino trilinear coupling
makes our vacuum unstable.
However, the vacuum stability bound in light mixed sneutrino WIMP scenarios
has been neglected in earlier works.
We compute the transition rate of our vacuum to a deeper one,
and show that the vacuum stability bound is not severe.
Although experimental constraints are very tight,
there are some regions where mixed sneutrino WIMP scenarios are viable.
We show that dark matter allowed regions can be examined by precisely measuring
the invisible decay rate of the observed Higgs boson
at future linear colliders.

The organization of this paper is as follows:
In Sec.II, the model of the mixed sneutrino dark matter is briefly reviewed.
Experimental constraints on the model are summarized in Sec.III\@.
In Sec.IV, the vacuum stability bound on our model is discussed.
Sec.V is devoted to a summary.

\section{Model}
\label{Sec:Model}

Here, we briefly review the mixed sneutrino model with lepton number
conservation, which is proposed in \cite{ArkaniHamed:2000bq}.
In this model, in addition to the usual 
matter content of the Minimal Supersymmetric Standard Model (MSSM), three
generations of right-handed neutrinos $\nu_{Ri}$ (sneutrinos
$\tilde{\nu}_{Ri}$) are introduced.  Here, $i=1,2,3$ denote the
generation.  As a result, Dirac neutrino Yukawa interactions, soft
right-handed sneutrino mass terms and soft trilinear couplings among
the left-handed slepton doublet $\widetilde \ell_i$, $\widetilde \nu_{Ri}$ and
the Higgs doublet with hypercharge $Y=1/2$, $h_u$, which gives mass to
the up-type quarks and Dirac neutrinos are added to the
usual MSSM Lagrangian.  The newly introduced soft terms are given by
\begin{eqnarray}
  \Delta {\cal L}_{\rm soft} = m^2_{\widetilde N_i}  |\widetilde \nu_{Ri} |^2 +  
        A_{\tilde\nu_i} \widetilde \ell_i \widetilde \nu_{Ri}^* h_u + {\rm h.c.} \,,
\end{eqnarray}
where ${m}^2_{\widetilde{N_i}}$ are soft right-handed sneutrino mass parameters,
and $A_{\widetilde\nu_i}$ are trilinear sneutrino $A$-parameters.
In order to avoid lepton flavor violation, 
we have assumed that these soft parameters are diagonal in generation space. 
Majorana neutrino mass terms and corresponding 
right-handed sneutrino bilinear terms are prohibited due to
lepton number conservation.

Neglecting the contribution from the Dirac neutrino masses, 
the sneutrino mass matrix for one generation is written as
\begin{eqnarray}
{\cal M}^2_{\tilde\nu} =
 \left( 
\begin{array}{cc}
  {m}^2_{\widetilde{L}} +\frac{1}{2} m^2_Z \cos 2\beta  &  \frac{1}{\sqrt{2}} A_{\tilde\nu}\, v \sin\beta\\
  \frac{1}{\sqrt{2}}    A_{\tilde\nu}\, v \sin\beta&  {m}^2_{\widetilde{N}}
 \end{array}\right) \, ,
\label{eq:sneutrino_tree}
\end{eqnarray}
where ${m}^2_{\tilde{L}}$ is the soft mass parameter for the
left-handed slepton doublet.  The sum of the squares of the vacuum
expectation values and the ratio of the vacuum expectation values are
given by $v^2=v_1^2+v_2^2=(246~{\rm GeV})^2$ and $\tan\beta=v_2/v_1$,
respectively. Here, $v_1$ ($v_2$) is the vacuum expectation value of
the Higgs doublet with hypercharge $Y=-1/2$ ($Y=1/2$).
In this model, the $A_{\tilde\nu}$ is not suppressed by the smallness
of the corresponding neutrino Yukawa coupling, but is
of the order of other soft parameters.
This large $A_{\tilde\nu}$ parameter gives a large mixing between the 
left-handed and right-handed sneutrinos, 
\begin{eqnarray}
 \tilde\nu_1 = \cos\theta_{\tilde\nu} \, \tilde\nu_R - \sin\theta_{\tilde\nu}\,  \tilde\nu_L \,, \quad
 \tilde\nu_2 = \sin\theta_{\tilde\nu} \, \tilde\nu_R + \cos\theta_{\tilde\nu}\,  \tilde\nu_L \,, 
\end{eqnarray}
with $m_{\tilde\nu_1} < m_{\tilde\nu_2}$, and the sneutrino 
mixing angle $\theta_{\tilde\nu}$ is given by
\begin{equation}
  \sin 2 \theta_{\tilde\nu} = \left( 
   \frac{\sqrt{2}A_{\tilde\nu}\, v\sin\beta }{ {m}^2_{\tilde\nu_2}-{m}^2_{\tilde\nu_1} } \right) \,.
\end{equation}

It should be emphasized that the couplings of the lighter sneutrino
to the $Z$-boson, the Higgs boson and neutralinos are suppressed by a power of the small mixing angle $\theta$,
compared to those of the MSSM left-handed sneutrinos.
The smallness of the sneutrino interactions plays an important role
in satisfying experimental constraints as discussed in the next section.
The Feynman rules for such sneutrino interactions are given by
\begin{eqnarray}
Z^{\mu} \tilde{\nu}^*_1(p') \tilde{\nu}_1(p) \ &:&
-i \frac{e}{\sin 2 \theta_W} (p+p')^{\mu}\sin^2\theta_{\tilde{\nu}}\, ,
\nonumber \\
h\tilde{\nu}^*_1\tilde{\nu}_1\ &:&
i e m_Z \frac{\sin(\alpha+\beta)}{\sin 2 \theta_W}
	\sin^2 \theta_{\tilde{\nu}}
	+i
\frac{2\cos \alpha}{v\sin \beta} \sin^2 \theta_{\tilde{\nu}} 
\cos^2 \theta_{\tilde{\nu}} ( m_{\tilde{\nu}_2}^2 - m_{\tilde{\nu}_1}^2)\, ,
\nonumber \\
\tilde{\nu}\overline{\nu_1}\widetilde{\chi}^0_i\ &:&
\frac{-ig}{2\sqrt{2} \sin 2\theta_W} (\cos\theta_W N_{i2} - \sin\theta_W N_{i1}) 
		\sin \theta_{\tilde{\nu}}  (1- \gamma_5)\, ,
\end{eqnarray}
where $e$ is the electric charge, $g$ the $SU(2)_L$ coupling constant,
$m_Z$ the $Z$-boson mass and $\theta_W$ the Weinberg angle. 
As for SUSY parameters, $\alpha$ is the Higgs mixing angle, and the matrix
$N_{ij}$ diagonalizes the neutralino mass matrix.

In the rest of this paper, for simplicity, we focus on the cases where
the lighter of the tau sneutrinos is a GeV-mass thermal WIMP
candidate.  We assume that the lighter sneutrinos of the first two
generations are too heavy to affect experimental constraints on such
GeV-mass tau sneutrino WIMP scenarios.

\section{Experimental constraints}
\label{Sec:Constraints}

Thermal WIMP candidates have been extensively tested through many experiments.
In particular, if the WIMP is lighter than half of the mass of the
Higgs boson and interacts with the Higgs boson, such light WIMP models
can be probed also through searches for the invisible decay of the Higgs boson.
We list relevant experimental constraints imposed on light tau sneutrino WIMP scenarios
in Table \ref{tab:expresulte}, and comment on the constraints below.
\begin{table}
\begin{center}
\caption{Observables and experimental constraints.}
\begin{tabular}{|c||c|c|}
	\hline
	Observable & Experimental result\\ \hline \hline
	$\Omega h^2$ & $0.1196 \pm 0.0062\ (95\%\ \mathrm{CL})$ \cite{Planck:2013pxb}\\  \hline
		$\sigma_{\rm N}^{\rm SI}$ & $(m_{\rm DM},\ \sigma_{\rm N}^{\rm SI})$ constraints \\
& from LUX \cite{Akerib:2013tjd} and SuperCDMS \cite{Agnese:2014aze} \\ \hline
	$\sigma_{\rm ann} v$  & $(m_{\rm DM},\ \sigma_{\rm ann}v)$ constraint \\
	& from FermiLAT \cite{Ackermann:2013yva} \\ \hline
	$\Delta \Gamma (Z \rightarrow \mathrm{inv.} )$ & $< 2.0\ \mathrm{MeV} \ (95\%\ \mathrm{CL})$
		 \cite{ALEPH:2005ab} \\ \hline
	$\mathrm{Br}(h \rightarrow \mathrm{inv.} )$ & $< 0.29 \ (95\%\ \mathrm{CL})$ \cite{ATLAS-CONF-2015-004} \\ \hline
	$m_{\tilde{\tau}_R}$ & $> 90.6\ \mathrm{GeV} \ (95\%\ \mathrm{CL})$ \cite{Aad:2014yka} \\ \hline
	$ m_{\widetilde{\chi}^{\pm}_1}$ & $> 420\ \mathrm{GeV} \ (95\%\ \mathrm{CL})$ \cite{Aad:2014yka} \\ \hline
	$ m_{\tilde{g}}$ & $> 1.4 \ \mathrm{TeV} \ (95\%\ \mathrm{CL})$
		\cite{Aad:2014lra, Chatrchyan:2014lfa} \\ \hline
\end{tabular}
\label{tab:expresulte}
	\end{center}
\end{table}

In general, dark matter candidates must be consistent with the upper limit of the 
dark matter relic density \cite{Planck:2013pxb}.
In our model, if the mass of the sneutrino WIMP is less than $10\
\mathrm{GeV}$, sneutrinos tend to annihilate into neutrinos via
neutralino exchange.  For $|M_{\widetilde{B}}| \ll |M_{\widetilde{W}}| \simeq |\mu|$, the lightest
neutralino is bino-like, and 
the thermal average of the sneutrino
annihilation cross section is given by 
\begin{equation}
\langle \sigma_{\rm ann} v \rangle 
=  \frac{\pi \alpha_{\rm em}^2 \sin^4 \theta_{\tilde{\nu}} }{256 \pi \sin^4 \theta_W \cos^4 \theta_W m^2_{\tilde{\chi}^0_1}}
\left(1- \frac{m_{\tilde{\nu}_1}^2}{m_{\tilde{\chi}^0_1}^2} \right)^2\, .
\label{eq:annihilation via neutralino}
\end{equation} 
The resulting thermal relic abundance of the sneutrino is approximately
\begin{equation}
	\Omega h^2 \sim 0.1 \times 
	\left ( \frac{\sin \theta_{\tilde{\nu}}}{0.1} \right)^{-4}
	\left( \frac{m_{\tilde{\chi}^0_1}}{1\ \mathrm{GeV}} \right)^2\, .
\label{eq:rough omega}
\end{equation}
Therefore, when the sneutrino mixing angle is 
as small as 0.1,
the relic abundance constraint requires the mass of bino-like neutralino
to be as small as ${\cal O}(1)$ GeV\@.
From this observation,
we concentrate on the cases where both the lightest tau sneutrino mass
and the bino-like neutralino mass are of the order of GeV\@.
Such a possibility has been overlooked in earlier works.

Next, let us discuss constraints from direct detection of dark matter.
For GeV-mass dark matter, the spin-independent scattering cross
section is limited by the LUX and the SuperCDMS experiments
\cite{Akerib:2013tjd,Agnese:2014aze}.  In our model, the scattering of
sneutrinos on nucleons occurs spin-independently via $Z$-boson or
Higgs boson exchange.  Since the $Z$-boson coupling to the sneutrino
dark matter candidate is suppressed by the square of the small mixing
angle $\theta_{\tilde \nu}^2$ compared to that to the MSSM left-handed
sneutrino, the resulting scattering cross section falls below its
experimental limit.  On the other hand, the coupling of the Higgs
boson to the sneutrino is proportional to the large $A$-term.  In the
nucleon scattering cross section, the ratio of the Higgs boson
exchange contribution to the $Z$-boson counterpart is proportional to
$m_{\tilde{\nu}_1}^{-2}$.  Actually, the amplitude of the scattering
via the Higgs boson is dominant over the one via the $Z$-boson for
$m_{\tilde{\nu}_1}^{} \sim {\cal O}(1)~{\rm GeV}$
\cite{Belanger:2010cd}.  The cross section of the scattering of the
dark matter and nucleon is given by:
\begin{equation}
	\sigma^{\rm SI}_{\rm N} 
	= \frac{4 \mu_{\chi} }{\pi}
		\frac{ ( Z f_p + (A-Z) f_n)^2 }{A^2},
\end{equation}
where $\mu_{\chi}$ is the sneutrino-nucleon reduced mass,
$A$ is the mass number, $Z$ is the atomic number 
and $f_p\ (f_n)$ is the amplitude for the proton (neutron).

As for indirect detection of dark matter, we impose the bound obtained
by the FermiLAT experiment on the annihilation cross section of the
sneutrino dark matter \cite{Ackermann:2013yva}.  In our model,
however, we have found that the constraint by the indirect detection
is not serious for GeV-mass sneutrino dark matter.

Let us turn to constraints from collider experiments.
The upper bound of the invisible decay of the $Z$-boson is obtained
at the LEP \cite{ALEPH:2005ab}:
\begin{eqnarray}
 \Delta \Gamma(Z \to \mathrm{inv.}) <  2.0\ \mathrm{MeV}\ (95\%\ \mathrm{CL}).
\end{eqnarray}
In our model, the $Z$-boson tends to decay invisibly 
to a lighter mixed sneutrino pair or a lightest neutralino pair.
The invisible decay width of the $Z$-boson to a pair of sneutrinos is 
proportional to the sneutrino mixing angle:
\begin{equation}
\Gamma (Z \rightarrow  \tilde{\nu}^*_1 \tilde{\nu}_1)
	= \Gamma (Z \rightarrow \bar{\nu} \nu)
		\frac{\sin^4 \theta_{\tilde{\nu}}}{2}
		\left(1-\frac{4m_{\tilde{\nu}_1}^2}{m_Z^2}\right)^{3/2},
\end{equation}
where $\Gamma (Z \rightarrow \bar{\nu} \nu)$ denotes the decay width of $Z$ boson to a pair of neutrinos:
\begin{equation}
\Gamma (Z \rightarrow \bar{\nu} \nu)
	=\frac{g^2}{96\pi \cos^2\theta_W} m_Z
	= 167\ \mathrm{MeV}.
\end{equation}
Therefore, the sneutrino mixing angle is constrained by
the result on the $Z$-boson invisible decay width.

Let us discuss experimental constraints on the Higgs boson invisible decay.
The branching ratio of the Higgs invisible decay is 
constrained directly through the searches for $Zh \rightarrow ll + E^{\rm miss}_{T}$
\cite{Aad:2014iia, Chatrchyan:2014tja},
and indirectly by the best-fit analysis using the combination of all channels of the Higgs boson decay
\cite{ATLAS-CONF-2015-004}.
Here, we employ the results of the best-fit constraint.
In our model, the decay width of the Higgs boson to a pair of the lighter mixed sneutrino 
is proportional to the sneutrino mixing angle $\sin^4 \theta_{\tilde{\nu}}$:
\begin{eqnarray}
\Gamma(h\rightarrow \tilde{\nu}_1 \tilde{\nu}_1^*)
	&=& \frac{ \sin^4 \theta_{\tilde{\nu}} }{16\pi m_h}
		\sqrt{ 1- \frac{4m_{\tilde{\nu}_1}^2}{m_{h}^2}} 
		\left| 
			 e m_Z \frac{\sin(\alpha+\beta)}{\sin 2 \theta_W}
			+\frac{2\cos \alpha}{v\sin \beta} \cos^2 \theta_{\tilde{\nu}} 
			( m_{\tilde{\nu}_2}^2 - m_{\tilde{\nu}_1}^2)
		\right|^2.
\end{eqnarray}
The Higgs boson can decay invisibly also to the lightest neutralino.
Such a decay mode is associated with the higgsino component of the lightest neutralino. When the $\mu$-parameter is much larger than the bino mass, the contribution from this invisible decay mode to the Higgs invisible decay is much smaller than 
the sneutrino pair channel.

We mention experimental constraints on the masses of electroweak
superparticles.  The pair production of sparticles is searched for at
the LEP, and the null results constrain the masses of the right-handed
sleptons, and the lightest chargino as shown in \cite{LEPresults}.
The LHC experiments also search for the pair productions of the
sleptons and the charginos \cite{Aad:2014vma,
  Aad:2014yka,Khachatryan:2014qwa}.  Such pair productions are
characterized by the signals for two leptons.  In addition, the
searches for the pair production of the lightest chargino and the
next-to-lightest neutralino impose the chargino mass limit more
strongly than the results of the chargino pair production.  In the
MSSM, the signal of the lightest chargino (next-to-lightest
neutralino) is characterized by a lepton (two leptons).  Therefore,
the chargino neutralino pair production is associated with the signal
of three leptons.  In our tau sneutrino WIMP model, the lightest
chargino dominantly decays to a tau with missing energy.  The modes
containing two taus account for half of the next-to-lightest
neutralino decay width, and most of the other half is converted to
missing energy without a charged track.  We use the constraints on the
lightest chargino mass by the searches for two or three taus.  In this
scenario, the mass of the lightest neutralino is close to that of the
LSP, and thus the lightest neutralino is long-lived and produces
displaced vertices in detectors.  Since the lightest neutralino decays
exclusively into a tau sneutrino and a tau neutrino, signatures of the
displaced vertices are invisible.  The search for the strong
production of sparticles in multi-$b$-jets final states constrains the
gluino mass \cite{Aad:2014lra, Chatrchyan:2014lfa}.

Finally, we comment on mono-photon searches at the LEP2 and LHC
experiments.  The LEP2 limit $\sigma(e^+ e^- \to \gamma + {\rm inv.})
< 15~{\rm pb}$ \cite{Achard:2003tx} does not place severe constraints
on the GeV-mass mixed sneutrino WIMP scenario \cite{Belanger:2010cd}.
In our model, the processes $q \bar{q} \to \gamma \widetilde{\nu}_1^{}
\widetilde{\nu}_1^*$ and $q \bar{q} \to\gamma \widetilde{\nu}_2^{}
\widetilde{\nu}_2^*$ mediated by the $Z$-boson give rise to events
with a mono-photon and missing transverse energy searched for at the
LHC~\cite{Aad:2014tda}.  The current LHC upper limit on the product of the vector boson
coupling to quarks and that to invisible particles is at most of the
order of unity, and thus not serious.

\section{Vacuum (meta-)stability bounds}
\label{Sec:Vacuum}

In the MSSM,
a large trilinear soft supersymmetry breaking term is known to
cause a minimum deeper than the Standard-Model-like (SML) vacuum
\cite{ccb}.
In our scenario, this is bound to be the case since
the neutrino masses are attributed to their small Yukawa couplings.
This is easy to see by tracing the scalar potential along the $D$-flat
direction,
\begin{equation}
 \label{eq:CCB direction}
 |h_u^0| = |\tilde{\nu}_L| = |\tilde{\nu}_R| = a,
\end{equation}
which leads to the lowest energy,
\begin{equation}
 \label{eq:VCCBmin}
 V_\mathrm{L.E.} =
 ( m^2_{h_u} + |\mu|^2 + m^2_{\widetilde{L}} + m^2_{\widetilde{N}} ) \, a^2
 - 2 |A_{\tilde{\nu}}| \, a^3 + 3 \lambda_\nu^2 \, a^4 ,
\end{equation}
where $\lambda_\nu$ is the neutrino Yukawa coupling.
One finds that $V_\mathrm{L.E.} < 0$ for some $a$ unless
the sneutrino trilinear coupling fulfils the inequality,
\begin{equation}
 |A_{\tilde{\nu}}|^2 \le
 3 ( m^2_{h_u} + |\mu|^2 + m^2_{\widetilde{L}} + m^2_{\widetilde{N}} )
 \, \lambda_\nu^2 ,
\end{equation}
which is the sneutrino-sector counterpart of the ``traditional'' bound
on $|A_{\tilde{t}}|$ from Charge-and-Color-Breaking (CCB) minima
\cite{ccb}.
This can be re-expressed in terms of
the sneutrino mass eigenvalues and mixing angle like
\begin{equation}
 \sin 2 \theta_{\tilde{\nu}} \le \frac{\sqrt{3} \, m_\nu
 ( m^2_{h_u} + |\mu|^2 + m^2_{\widetilde{L}} + m^2_{\widetilde{N}} )^{1/2}}
 {m^2_{\tilde{\nu}_2} - m^2_{\tilde{\nu}_1}} .
\end{equation}
This means that $\theta_{\tilde{\nu}} \gtrsim 2\times 10^{-12}$
implies a lepton-number breaking global minimum,
if one assumes that $m_\nu \sim 1\ \mathrm{eV}$ and
all the other mass parameters above
are around $100\ \mathrm{GeV}$.
Therefore, in the range of $\theta_{\tilde{\nu}}$
required by a viable relic density of light sneutrino DM,
the SML vacuum is inevitably a local minimum with a finite lifetime.

Given the low value of $m^2_{\widetilde{L}}$,
our model can also develop an unbounded-from-below (UFB) direction,
if $m^2_{h_u} + m^2_{\widetilde{L}} < 0$ \cite{Komatsu:1988mt}.
However, we shall not consider this direction for the reason
to be explained below.

In order to judge whether the global minimum invalidates
this model or not, one would need to consider two aspects:
the cosmological history of the vacuum, and
the lifetime of the current SML vacuum.
Regarding the former, one could argue that
inflation-induced scalar masses might have brought the Universe
to the SML vacuum \cite{inflationary mass}.
The latter then becomes the remaining criterion.

Employing a semiclassical approximation \cite{tunnelling},
one can express the false vacuum decay rate per unit volume in the form,
\begin{equation}
 \Gamma/V = A\, \exp( -S [\overline{\phi}] ) ,
\end{equation}
where $A$ is a prefactor which we set to $(100\ \mathrm{GeV})^4$
on dimensional grounds, $S$ is the Euclidean action, and
$\overline{\phi}$ is an O(4)-symmetric \cite{Coleman:1977th}
stationary point of
\begin{equation}
 S[\phi(\rho)] = 2 \pi^2 \int_0^\infty d\rho \rho^3
 \left[ \left|\frac{d\phi}{d\rho}\right|^2 + V(\phi) \right] .
\end{equation}
The ``bounce'' $\overline{\phi}(\rho)$
shall obey the boundary conditions,
\begin{equation}
 \overline{\phi}(\rho \rightarrow \infty) = \phi_+, \quad
 \frac{d\overline{\phi}}{d\rho} (\rho=0) = 0 ,
\end{equation}
where $\phi_+$ denotes the false vacuum.
The criterion for admitting a parameter set shall be
$S[\overline{\phi}] > 400$
which is the requirement that
the lifetime of the observable spatial volume at the SML vacuum
be longer than the age of the Universe \cite{Claudson:1983et}.

To obtain the bounce configuration $\overline{\phi}(\rho)$,
we use the numerical method described in Ref.~\cite{Park:2010rh}
which works even for a scalar potential with a distant or non-existent
global minimum.
For a fast computation,
we restrict the set $\phi$ of scalar fields to
$\{ h_d^0, h_u^0, \tilde{\nu}_L, \tilde{\nu}_R \}$.
The other scalars are assumed to be zero along the bounce, but
we do not impose on it
any extra constraint such as $D$-flatness.
In view of the shape of the potential,
this should not preclude a tunnelling path possibly with a lower $S$.
This field restriction excludes the UFB-3 direction
in Ref.~\cite{Casas:1995pd}
which is a generalization of the aforementioned UFB direction
\cite{Komatsu:1988mt},
since these directions would require two more non-vanishing scalar fields,
e.g.\
a pair of down-type squarks or sleptons of a different generation from
the sneutrino generation.
However, such UFB paths contain intervals with non-vanishing $D$-terms
which form high potential barriers.
Therefore, contributions from the UFB paths to $\Gamma/V$ would be
highly suppressed compared to that from a path throughout which
the $D$-terms are negligible.

As a way to check the validity of our program,
we compared its value of $S$ to
that from CosmoTransitions \cite{Wainwright:2011kj},
using the two-scalar toy model included in the package.

With the tree-level potential,
plus a term proportional to $|h_u|^4$ for fitting the measured Higgs mass
(see e.g.\ \cite{Hisano:2010re}),
one can determine the tunnelling rate by fixing
$m_{\tilde{\nu}_1}$, $m_{\tilde{\nu}_2}$, $\theta_{\tilde{\nu}}$,
$\mu$, $\tan\beta$, and $M_A$,
the last of which is the $CP$-odd Higgs mass.
To obtain the vacuum lifetime bound on $\theta_{\tilde{\nu}}$,
we set 
$M_A = 400\ \mathrm{GeV}$,
while we choose the other parameters as in Table~\ref{tab:inputparameter}.
Note that the tunnelling rate is insensitive to $M_A$
for $\tan\beta \gtrsim 10$ since
the $CP$-odd Higgs as well as the other extra Higgses belong
mostly to $h_d$
whose components remain to be small along the bounce.
(A similar discussion about the irrelevance of $M_A$ to the bounds on
flavour-violating up-type trilinears is found in Ref.~\cite{Park:2010wf}.)

The overall conclusion from
the numerical computation 
turns out to be that
the vacuum longevity constraint on $\theta_{\tilde{\nu}}$ is
so loose that it allows the entire range of $\theta_{\tilde{\nu}}$
limited by $Z \rightarrow \mathrm{inv.}$ and $h \rightarrow \mathrm{inv}$.
For instance, any $\theta_{\tilde{\nu}} \le 0.52$ is safe
from rapid bubble nucleation for $m_{\tilde{\nu}_1} = 0.1\ \mathrm{GeV}$.
Even larger $\theta_{\tilde{\nu}}$ is allowed for higher
$m_{\tilde{\nu}_1}$, since $A_{\tilde{\nu}}$ which triggers the tunnelling
is proportional to $m^2_{\tilde{\nu}_2} - m^2_{\tilde{\nu}_1}$.
This trend continues up to the point
$m_{\tilde{\nu}_1} \simeq 10\ \mathrm{GeV}$ where
the upper bound disappears, i.e.\ $S > 400 $ for any $\theta_{\tilde{\nu}}$.

Apart from the above constraint at zero temperature,
the vacuum stability at high temperatures is known to exclude
potentially more parameter volume \cite{Kawasaki:2000ye,Camargo-Molina:2014pwa}.
For instance,
Fig.~1 of Ref.~\cite{Camargo-Molina:2014pwa} shows that
thermal effects might decrease the bound
on the stop trilinear by about 20\%,
in the parameter space considered therein.
Naively scaling the limit on the sneutrino trilinear by the same factor,
one might expect to be safe provided that $\theta_{\tilde{\nu}} \le 0.38$,
which is still far above the collider bounds.
We leave an explicit check of this point as a future work.

\section{Results}
\label{Sec:Results}

We analyze the GeV-mass region of the thermal mixed sneutrino dark matter scenarios.
Since direct detections have an energy threshold, in general it is difficult to detect the GeV-mass WIMP directly.
However, if the GeV-mass WIMP interacts with the Higgs boson,
the search for the Higgs boson invisible decay can constrain the parameter space of
the GeV-mass WIMP\@.
In the thermal light mixed sneutrino scenarios, 
the Higgs invisible decay imposes the upper limit on the sneutrino mixing angle.
The small mixing angle of the sneutrino requires that 
the mass of the lightest neutralino is of the order of $1\ \mathrm{GeV}$
(see Eq.(\ref{eq:rough omega})).
On the other hand, the GUT relation $6M_{ \widetilde{B}} = 3M_{ \widetilde{W}} = M_{\tilde{g}}$,
which is assumed in earlier works, and the experimental constraints on 
the gluino mass \cite{Aad:2014lra, Chatrchyan:2014lfa}
require the lightest neutralino mass to be $\mathcal{O}(100\ \mathrm{GeV})$.
Thus, we relax the GUT relation and focus on the GeV-mass region of 
the thermal mixed sneutrino dark matter and the lightest neutralino.

\begin{table}[t]
	\begin{center}
	\caption{Parameters and reference values/scan bounds. 
	}
	\label{tab:inputparameter}
	\begin{tabular}{|c||c|c|}
	\hline
	Parameter & Reference value/Scan bound\\ \hline \hline
	$ \mu $ & $500\ \mathrm{GeV} $ \\ \hline
	$ \tan \beta$ & $10 $ \\ \hline
	$m_{\tilde{\nu}_2}$ & $125\ \mathrm{GeV} $ \\ \hline
	$m_{\tilde{\tau}_R}$ & $120\ \mathrm{GeV}$ \\ \hline
	$M_{\widetilde{W}}$ & $500\ \mathrm{GeV}$ \\ \hline
	$m_{\tilde{\nu}_1}$ & $[0.1\ \mathrm{GeV},\ 10\ \mathrm{GeV}] $ \\ \hline
	$\sin \theta_{\tilde{\nu}}$& $[0.01,\ 0.3] $ \\ \hline
	$M_{\widetilde{B}}$ & $[0.1 \ \mathrm{GeV} ,\ 20\
	\mathrm{GeV} ] $ \\ \hline
	\end{tabular}
	\end{center}
\end{table}

Model parameters and their reference values or scan bounds are
summarized in Table \ref{tab:inputparameter}.  We searched for the
parameter set which minimizes the branching ratio of the Higgs invisible decay
for a given sneutrino dark matter mass.  We comment on the
choices of the reference values below.  To reduce the higgsino
component of the lightest neutralino, the $\mu$-parameter is set to as
large as $500$ GeV\@.  Then, the decay width of the Higgs boson to a
pair of lightest neutralinos is adequately suppressed.  The reference
value of $\tan\beta=10$ is chosen to obtain a 125 GeV Higgs boson for
less hierarchical superparticle mass spectra.  Our results do not
strongly depend on the choice of $\tan\beta$ except for the MSSM Higgs
boson properties.  The heavier sneutrino mass should not be smaller
than the Higgs boson mass in order to suppress the decay width of the
Higgs boson to a pair of the lighter and heavier sneutrinos.  On the
other hand, since the sneutrino $A$-term, which triggers the false
vacuum decay, is proportional to
$m^2_{\tilde{\nu}_2}-m^2_{\tilde{\nu}_1}$, the heavier sneutrino
should be light enough. Therefore, we set $m_{{\tilde
    \nu}_2}^{}=m_h=125$ GeV\@.  We choose possible smallest values for
the right-handed stau mass and the wino mass in the light of the LHC
results about the two and three tau searches \cite{Aad:2014yka}.  The
colored superparticles as well as first two generations of sleptons
are assumed to be too heavy to affect our numerical results.

For our numerical computations of dark matter properties, we have
implemented SUSY model files containing right-handed (s)neutrino
interactions into the public code {\tt micromegas\,3.2}
\cite{Belanger:2013oya}.  The model files are generated with the help of
the Feynman rules generation tool {\tt LanHEP\,3.1.8}
\cite{Semenov:2010qt}.

\begin{figure}[t]
\includegraphics{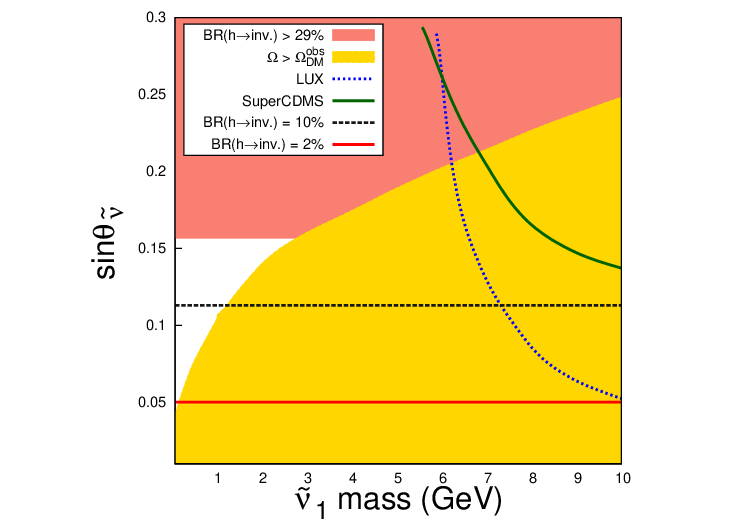}
\caption{\footnotesize The results of our parameter scan for light
  mixed sneutrino dark matter scenarios in the $(m_{{\tilde \nu}_1},
  \sin \theta_{\tilde{\nu}})$ plane.  The yellow (light-gray) and pink
  (dark-gray) regions are ruled out by the results of the relic
  abundance \cite{Planck:2013pxb} and the Higgs boson invisible decay
  \cite{ATLAS-CONF-2015-004}, respectively.  We also show the upper
  limits of the spin-independent elastic WIMP-nucleon cross section by
  the LUX (blue dotted line) \cite{Akerib:2013tjd} and the SuperCDMS
  (dark-green line) \cite{Agnese:2014aze}.  The black dashed (red
  solid) line corresponds to the Higgs boson invisible decay branching
  fraction of $10\%$ ($2\%$).  }
\label{fig:general1g}
\end{figure}

Fig.~\ref{fig:general1g} shows the results of our parameter scan in the 
$(m_{\tilde{\nu}_1}, \sin \theta_{\tilde{\nu}})$ plane.
In the yellow (light-grey) region, the relic density of the sneutrino
is larger than the observed dark matter relic density obtained by the
Planck collaboration \cite{Planck:2013pxb}.  The pink (dark-grey)
region is excluded by the Higgs boson invisible decay searches at
ATLAS \cite{ATLAS-CONF-2015-004}.  In the allowed region, the spin
independent cross section is of the order of $10^{-42}\
\mathrm{cm}^2$. The constraints on dark matter direct detection rates
from LUX (blue dotted line) \cite{Akerib:2013tjd} and SuperCDMS
(green solid line) \cite{Agnese:2014aze} are not serious for such
light sneutrino masses as shown in Fig.~\ref{fig:general1g}.

It should be emphasized that 
Higgs boson invisible decay
searches at future collider experiments will give a stronger constraint on 
such light mixed sneutrino WIMP scenarios.
If the Higgs boson invisible decay branching ratio is constrained to $10\%$
($2\%$),
the sneutrino mixing angle $\sin \theta_{\tilde{\nu}}$ must be smaller 
than $0.12$ ($0.05$).
The expectations of the ATLAS and CMS collaborations on the Higgs boson invisible decay are 
${\rm Br} < 8.0\%$ ($95\%~{\rm CL}$) \cite{ATL-PHYS-PUB-2013-014}
and ${\rm Br} < 6.4\%$ ($95\%~{\rm CL}$) \cite{CMS:2013xfa}, respectively,
at the LHC high-luminosity program with the center-of-mass energy of $\sqrt{s} = 14~{\rm TeV}$ and the luminosity of $L = 3000~{\rm fb^{-1}}$.
The planned International Linear Collider (ILC) is capable of
measuring the Higgs boson invisible branching ratio accurately
\cite{Asner:2013psa,Baer:2013cma}.
Using the polarization configuration $(P_{e^-}, P_{e^+}) = (+80\%,
-30\%)$ with $\sqrt{s}=250~{\rm GeV}$ and $L = 250~{\rm fb^{-1}}$, the
upper limit will reach ${\rm Br}(h\to {\rm inv.})<0.69\%$ ($95\%~{\rm
  CL}$) \cite{Ishikawa}.
This means that the ILC is capable of excluding mixed sneutrino WIMP
scenarios for $0.1~{\rm GeV} < m_{\tilde{\nu}_1} < m_h/2$.

\section{Conclusions}

In supersymmetric models with Dirac neutrino masses where 
soft breaking trilinear sneutrino interactions
are not suppressed by small neutrino Yukawa couplings,
the lightest mixed sneutrino is one of the viable thermal WIMP 
candidates due to the non-negligible
mixings between the left- and right-handed states.
We have focused on the cases where the lighter of the mixed tau 
sneutrinos
is a WIMP with mass of the order of $1$ GeV, and
investigated phenomenological constraints on such scenarios.
We have shown that 
if the mass of the bino-like neutralino is also of the order
of GeV, the dark matter relic abundance can be explained
while adequately suppressing the invisible Higgs boson decay rate.
This situation could be realized by relaxing
gaugino mass universality which, if retained,
would have disabled our scenario
because of the severe gluino mass bound obtained at the LHC\@.

Special attention has been paid to the vacuum stability bound.
The large trilinear soft breaking sneutrino interaction also
makes a lepton number violating vacuum deeper than the MSSM-like vacuum.
We have computed the relevant Euclidean action,
and shown that the lifetime of the Universe
in the current phase is long enough
in the allowed regions where the dark matter and Higgs invisible decay 
constraints are satisfied.

Although dark matter direct detections cannot give stringent constraints
on such a low mass WIMP, we have shown that the ILC
has the ability to explore the allowed region through the Higgs invisible 
decay
search if the mass of the mixed tau sneutrino is larger than 0.1 GeV\@.
Such light mixed sneutrino scenarios are good examples to show 
that future linear colliders can explore model parameter regions
which other experiments cannot probe.

\begin{acknowledgments}
We thank
Junji Hisano and Florian Staub
for bringing Refs.~\cite{Kawasaki:2000ye} and~\cite{Camargo-Molina:2014pwa}
to our attention, respectively, along with the valuable comments.
We also thank Akimasa Ishikawa for useful discussions
and bringing Ref.~\cite{Ishikawa} to our attention.
The work of M.K. was supported in part by Grant-in-Aid for Scientific Research, 
Japan Society for the Promotion of Science (JSPS), No.\ 26104702.
J.P. acknowledges support from the MEC and FEDER (EC) Grants
FPA2011--23596 and the Generalitat Valenciana under grant PROMETEOII/2013/017.

\end{acknowledgments}


\begin{thebibliography}{99}

\bibitem{LHC} 
 G.~Aad {\it et al.}  [ATLAS Collaboration],
 Phys.\ Lett.\ B {\bf 716}, 1 (2012), 
 [arXiv:1207.7214]; 
 S.~Chatrchyan {\it et al.}  [CMS Collaboration],
 Phys.\ Lett.\ B {\bf 716}, 30 (2012), 
 [arXiv:1207.7235]. 


\bibitem{SK}
 Y.~Fukuda {\it et al.} [Super-Kamioande Collaboration], 
 Phys.\ Rev.\ Lett.\  {\bf 81}, 1562 (1998), 
 [hep-ex/9807003].

\bibitem{WMAP}
 G. Hinshaw {\it et al.} [WMAP Collaboration],
 Astrophys.\ J.\ Suppl.\ {\bf 208}, 19 (2013), 
 [arXiv:1212.5226]. 


\bibitem{Planck:2013pxb}
P.~A.~R.~Ade \textit{et al.} [Planck],
Astron. Astrophys. \textbf{571}, A16 (2014),
[arXiv:1303.5076]. 


 \bibitem{HallMoroi:1997}
 L. J. Hall, T. Moroi, and H. Murayama, 
 Phys.\ Lett.\ B {\bf 424}, 305 (1998), 
 [hep-ph/9712515].

\bibitem{ArkaniHamed:2000bq} 
 N.~Arkani-Hamed, L.~J.~Hall, H.~Murayama, D.~Tucker-Smith and N.~Weiner,
 Phys.\ Rev.\ D {\bf 64}, 115011 (2001),
 [hep-ph/0006312].

 \bibitem{BorzumatiNomura:2000}
 F. Borzumati and Y. Nomura, 
 Phys.\ Rev.\ D {\bf 64}, 053005 (2001), 
 [hep-ph/0007018].

\bibitem{Alan:2004}
 A. T. Alan and S. Sultansoy, 
 J.\ Phys.\ G{\bf 30}, 937 (2004),
 [hep-ph/0307143].

 \bibitem{AsakaMoroi:2005}
 T. Asaka, K. Ishiwata, and T. Moroi, 
 Rev.\ D{\bf 73}, 051301 (2006),
 [hep-ph/0512118].


\bibitem{Arina:2007}
 C. Arina and N. Fornengo, 
 JHEP {\bf 0711}, 029 (2007), 
 [arXiv:0709.4477].

\bibitem{Thomas:2008}
 Z. Thomas, D. Tucker-Smith, and N. Weiner, 
 Phys.\ Rev.\ D{\bf 77}, 115015 (2008), 
 [arXiv:0712.4146].


\bibitem{Deppisch}
 F. Deppisch and A. Pilaftsis, 
 JHEP {\bf 0810}, 080 (2008), 
 [arXiv:0808.0490].

\bibitem{CerdenoSeto:2009}
 D. G. Cerdeno and O. Seto, 
 JCAP {\bf 0908}, 032 (2009), 
 [arXiv:0903.4677].

\bibitem{Belanger:2010cd} 
 G.~Belanger, M.~Kakizaki, E.~K.~Park, S.~Kraml and A.~Pukhov,
 JCAP {\bf 1011}, 017 (2010),
 [arXiv:1008.0580]. 

\bibitem{Belanger:2011}
 G. Belanger, S. Kraml, and A. Lessa, 
 JHEP {\bf 1107}, 083 (2011), 
 [arXiv:1105.4878].

\bibitem{KhalilOkada:2011}
 S. Khalil, H. Okada, and T. Toma, 
 JHEP {\bf 1107}, 026 (2011), 
 [arXiv:1102.4249].

\bibitem{ChoiSeto:2012}
 K.-Y. Choi and O. Seto, 
 Phys.\ Rev.\ D{\bf 86}, 043515 (2012), 
 [arXiv:1205.3276].

\bibitem{Dumont:2012ee} 
 B.~Dumont, G.~Belanger, S.~Fichet, S.~Kraml and T.~Schwetz,
 JCAP {\bf 1209}, 013 (2012),
 [arXiv:1206.1521] 

\bibitem{Arina:2014}
 C. Arina and M. E. Cabrera, 
 JHEP {\bf 1404}, 100 (2014), 
 [arXiv:1311.6549].

\bibitem{Arina:2015uea} 
  C.~Arina, M.~E.~C.~Catalan, S.~Kraml, S.~Kulkarni and U.~Laa,
  arXiv:1503.02960 [hep-ph].

\bibitem{Akerib:2013tjd} 
 D.~S.~Akerib {\it et al.}  [LUX Collaboration],
 Phys.\ Rev.\ Lett.\  {\bf 112}, no.\ 9, 091303 (2014),
 [arXiv:1310.8214]. 

\bibitem{Agnese:2014aze} 
 R.~Agnese {\it et al.}  [SuperCDMS Collaboration],
 Phys.\ Rev.\ Lett.\  {\bf 112}, 241302 (2014),
 [arXiv:1402.7137]. 

\bibitem{Ackermann:2013yva} 
 M.~Ackermann {\it et al.}  [Fermi-LAT Collaboration],
 Phys.\ Rev.\ D {\bf 89}, 042001 (2014),
 [arXiv:1310.0828]. 

\bibitem{ALEPH:2005ab} 
 S.~Schael {\it et al.} 
 Phys.\ Rept.\  {\bf 427}, 257 (2006),
 [hep-ex/0509008].

\bibitem{ATLAS-CONF-2015-004} 
  The ATLAS collaboration,
Boson Fusion in $pp$ Collisions at $\sqrt{s}=8$ TeV using the ATLAS 
Detector at the LHC,''
  ATLAS-CONF-2015-004, ATLAS-COM-CONF-2015-004.


\bibitem{Aad:2014yka} 
 G.~Aad {\it et al.}  [ATLAS Collaboration],
 JHEP {\bf 1410}, 96 (2014),
 [arXiv:1407.0350]. 

\bibitem{Aad:2014lra} 
 G.~Aad {\it et al.}  [ATLAS Collaboration],
 JHEP {\bf 1410}, 24 (2014),
 [arXiv:1407.0600]. 

\bibitem{Chatrchyan:2014lfa} 
 S.~Chatrchyan {\it et al.}  [CMS Collaboration],
 JHEP {\bf 1406}, 055 (2014),
 [arXiv:1402.4770]. 



\bibitem{Aad:2014iia} 
 G.~Aad {\it et al.}  [ATLAS Collaboration],
 Phys.\ Rev.\ Lett.\  {\bf 112}, 201802 (2014),
 [arXiv:1402.3244]. 

\bibitem{Chatrchyan:2014tja} 
 S.~Chatrchyan {\it et al.}  [CMS Collaboration],
 Eur.\ Phys.\ J.\ C {\bf 74}, no.\ 8, 2980 (2014),
 [arXiv:1404.1344]. 

\bibitem{LEPresults}
 LEP2 SUSY Working Group [ALEPH, DELPHI, L3, OPAL Collaboration],
 \\http://lepsusy.web.cern.ch/lepsusy/.

\bibitem{Aad:2014vma} 
 G.~Aad {\it et al.}  [ATLAS Collaboration],
 JHEP {\bf 1405}, 071 (2014),
 [arXiv:1403.5294]. 

\bibitem{Khachatryan:2014qwa} 
 V.~Khachatryan {\it et al.}  [CMS Collaboration],
 Eur.\ Phys.\ J.\ C {\bf 74}, no.\ 9, 3036 (2014),
 [arXiv:1405.7570]. 

\bibitem{Achard:2003tx} 
  P.~Achard {\it et al.}  [L3 Collaboration],
  Phys.\ Lett.\ B {\bf 587}, 16 (2004)
  [hep-ex/0402002].


\bibitem{Aad:2014tda} 
  G.~Aad {\it et al.}  [ATLAS Collaboration],
  Phys.\ Rev.\ D {\bf 91}, no. 1, 012008 (2015)
  [arXiv:1411.1559 [hep-ex]].


\bibitem{ccb}
 J.~M.~Frere, D.~R.~T.~Jones and S.~Raby,
 Nucl.\ Phys.\  B {\bf 222}, 11 (1983);
 L.~Alvarez-Gaume, J.~Polchinski and M.~B.~Wise,
 Nucl.\ Phys.\  B {\bf 221}, 495 (1983);
 J.~P.~Derendinger and C.~A.~Savoy,
 Nucl.\ Phys.\  B {\bf 237}, 307 (1984);
 C.~Kounnas, A.~B.~Lahanas, D.~V.~Nanopoulos and M.~Quiros,
 Nucl.\ Phys.\  B {\bf 236}, 438 (1984).

\bibitem{Komatsu:1988mt}
 H.~Komatsu,
 Phys.\ Lett.\ B {\bf 215}, 323 (1988).

\bibitem{inflationary mass}
 M.~Dine, L.~Randall and S.~D.~Thomas,
 Phys.\ Rev.\ Lett.\  {\bf 75}, 398 (1995)
 [hep-ph/9503303];
 J.~R.~Ellis, J.~Giedt, O.~Lebedev, K.~Olive and M.~Srednicki,
 Phys.\ Rev.\  D {\bf 78}, 075006 (2008)
 [arXiv:0806.3648]. 

\bibitem{tunnelling}
 S.~R.~Coleman,
 Phys.\ Rev.\  D {\bf 15}, 2929 (1977)
 [Erratum-ibid.\  D {\bf 16}, 1248 (1977)];
 C.~G.~Callan, Jr.\ and S.~R.~Coleman,
 Phys.\ Rev.\  D {\bf 16}, 1762 (1977).

\bibitem{Coleman:1977th}
 S.~R.~Coleman, V.~Glaser and A.~Martin,
 Commun.\ Math.\ Phys.\  {\bf 58}, 211 (1978).

\bibitem{Claudson:1983et}
 M.~Claudson, L.~J.~Hall and I.~Hinchliffe,
 Nucl.\ Phys.\  B {\bf 228}, 501 (1983).

\bibitem{Park:2010rh}
 J.-h.~Park,
 JCAP {\bf 1102}, 023 (2011),
 [arXiv:1011.4936]. 

\bibitem{Casas:1995pd}
 J.~A.~Casas, A.~Lleyda and C.~Munoz,
 Nucl.\ Phys.\  B {\bf 471}, 3 (1996),
 [hep-ph/9507294].

\bibitem{Wainwright:2011kj}
 C.~L.~Wainwright,
 Comput.\ Phys.\ Commun.\  {\bf 183}, 2006 (2012),
 [arXiv:1109.4189]. 

\bibitem{Hisano:2010re}
 J.~Hisano and S.~Sugiyama,
 Phys.\ Lett.\ B {\bf 696}, 92 (2011)
  [Erratum-ibid.\ B {\bf 719}, 472 (2013)],
 [arXiv:1011.0260]. 

\bibitem{Park:2010wf}
 J.-h.~Park,
 Phys.\ Rev.\ D {\bf 83}, 055015 (2011),
 [arXiv:1011.4939]. 

\bibitem{Kawasaki:2000ye} 
 M.~Kawasaki, T.~Watari and T.~Yanagida,
 Phys.\ Rev.\ D {\bf 63}, 083510 (2001),
 [hep-ph/0010124].

\bibitem{Camargo-Molina:2014pwa} 
 J.~E.~Camargo-Molina, B.~Garbrecht, B.~O'Leary, W.~Porod and F.~Staub,
 Phys.\ Lett.\ B {\bf 737}, 156 (2014),
 [arXiv:1405.7376]. 

\bibitem{Belanger:2013oya} 
  G.~Belanger, F.~Boudjema, A.~Pukhov and A.~Semenov,
  Comput.\ Phys.\ Commun.\  {\bf 185}, 960 (2014)
  [arXiv:1305.0237 [hep-ph]].

\bibitem{Semenov:2010qt} 
  A.~Semenov,
  arXiv:1005.1909 [hep-ph].


\bibitem{ATL-PHYS-PUB-2013-014}
  ALTAS Collaboration, ATL-PHYS-PUB-2013-014.

\bibitem{CMS:2013xfa} 
  CMS Collaboration,
  arXiv:1307.7135.

\bibitem{Asner:2013psa} 
  D.~M.~Asner, T.~Barklow, C.~Calancha, K.~Fujii, N.~Graf, H.~E.~Haber, 
A.~Ishikawa and S.~Kanemura {\it et al.},
  arXiv:1310.0763 [hep-ph].

\bibitem{Baer:2013cma} 
  H.~Baer, T.~Barklow, K.~Fujii, Y.~Gao, A.~Hoang, S.~Kanemura, J.~List 
and H.~E.~Logan {\it et al.},
  arXiv:1306.6352 [hep-ph].

\bibitem{Ishikawa} 
  A.~Ishikawa,
  Talk at the 16th International Workshop on Future Linear Collider (LCWS14),
  Belgrade, Serbia, October 6-10, 2014.
\end{thebibliography}
\end{document}